\documentclass{moriond}

\usepackage{graphicx}
\usepackage{amsmath,amssymb}
\usepackage{xspace}

\usepackage{multicol,caption}

\newcommand{\pp}{\mbox{$pp$}\xspace}
\newcommand{\zz}{\mbox{$z_0\sin\theta$}}
\newcommand{\dz}{\mbox{$d_0$}}
\newcommand{\edz}{\mbox{$\sigma_{\dz}$}}
\newcommand{\ezz}{\mbox{$\sigma_{\zz}$}}
\newcommand{\akt}{\mbox{anti-$k_{t}$}\xspace}
\newcommand{\Raa}{\mbox{$R_{\mathrm{AA}}$}\xspace}
\newcommand{\Npart}{\mbox{$N_{\mathrm{part}}$}\xspace}
\newcommand{\avgNpart}{\mbox{$\langle \Npart \rangle$}\xspace}
\newcommand{\Taa}{\mbox{$T_{\mathrm{AA}}$}\xspace}
\newcommand{\avgTaa}{\mbox{$\langle \Taa \rangle$}\xspace}
\newcommand{\sqn}{\mbox{$\sqrt{s_{\mathrm{NN}}}$}\xspace}
\newcommand{\sqs}{\mbox{$\sqrt{s}$}\xspace}
\newcommand{\PbPb}{\mbox{Pb+Pb}\xspace}
\newcommand{\pT}{\mbox{$p_\mathrm{T}$}\xspace}
\newcommand{\TeV}{\mbox{Te\kern -0.1em V}\xspace}
\newcommand{\GeV}{\mbox{Ge\kern -0.1em V}\xspace}

\newcommand{\Photo}{}

\newenvironment{Figure}
  {\par\medskip\noindent\minipage{\linewidth}}
  {\endminipage\par\medskip}

\begin{document}

\title{Measurement of the charged-hadron spectra and nuclear modification factor in lead--lead collisions with the ATLAS detector}

\author{Petr Balek (for the ATLAS Collaboration)}
\address{IPNP, Faculty of Mathematics and Physics, Charles University in Prague, V Hole\v{s}ovi\v{c}k\'{a}ch 2,  180 00 Prague,  Czech Republic}

\maketitle

\begin{abstract}
The ATLAS experiment at the Large Hadron Collider (LHC) measures charged hadron spectra obtained in 2010 and 2011 lead--lead LHC data taking periods with total integrated statistics of 0.15\,nb${}^{-1}$. The results are compared to the \pp spectra of charged hadrons at the same centre-of-mass energy based on the data sample with integrated luminosity of 4.2\,pb${}^{-1}$ obtained by the ATLAS experiment in 2011 and 2013. This allows a detailed comparison of the two collision systems in a wide transverse momentum ($0.5<\pT<150$\,\GeV) and pseudorapidity \mbox{($|\eta|<2$)} ranges in different centrality intervals of Pb+Pb collision. The nuclear modification factor \Raa is presented in detail as a function of centrality, $\pT$ and $\eta$. It shows a distinct $\pT$-dependence with a pronounced minimum at about 7\,\GeV. Above 60\,\GeV, it is consistent with a flat, centrality-de\-pen\-dent, value within the uncertainties. The value is $0.55\pm0.01(stat.)\pm0.04(syst.)$ in the most central collisions. The \Raa is observed to be consistent with flat $|\eta|$ dependence over the whole transverse momentum range in all centrality classes.

\hspace*{1cm}

\noindent \textit{keywords}: lead--lead collision, nuclear modification factor
\end{abstract}

\begin{multicols}{2}

\section{Introduction}
The first results from the LHC experiments showed that jets emerging from the quark-gluon plasma, the hot and dense matter produced in the heavy-ion (HI) collisions, have lower energy than would be expected in the absence of medium effects~\cite{Aj_atlas,Aj_cms}. The measurement of jets by the ATLAS and ALICE experiments revealed that yield of jets at fixed transverse momentum ($\pT$), caused by the energy loss of the high-energy partons in the medium, is suppressed by a factor of two to four~\cite{atlas_jet_raa,alice_jet_raa}. Additional information about the energy-loss mechanism is obtained from the study of jet fragmentation functions that show a small enhancement at $\pT\gtrsim 30$\,\GeV~\cite{frag_atlas,frag_cms}.

Production of hadrons originating from jet fragmentations is expected to be modified as well, making charged hadrons usable in a study of the quark-gluon plasma. The results from the LHC experiments show that the suppression reaches a factor of seven~\cite{atlas_raa,cms_raa,alice_raa} at $\pT$ around 7\,\GeV for HI collisions at \mbox{$\sqn=2.76$\,\TeV.} Results from RHIC with $\sqn=200$\,\GeV show a suppression by a factor of five~\cite{brahms,phobos,star,phenix}.

\section{Analysis}
The suppression of hadron production in HI collisions is quantified by the nuclear modification factor ($\Raa$) defined as
\begin{equation}
\Raa = \frac{1}{\avgTaa} \frac{1/N_\mathrm{evt}~\mathrm{d}^2N_{\footnotesize\PbPb} / \mathrm{d}\eta \mathrm{d}\pT}{\mathrm{d}^2\mathrm{\sigma}_{\footnotesize\pp} / \mathrm{d}\eta \mathrm{d}\pT},
\end{equation}
\noindent where $\avgTaa$ is nuclear thickness function accounting for increased flux of partons per collision in \PbPb collisions, $N_\mathrm{evt}$ is the number of \PbPb events, $\mathrm{d}^2N_{\footnotesize\PbPb} / \mathrm{d}\eta \mathrm{d}\pT$ is the differential yield of charged particles in \PbPb collisions and $\mathrm{d}^2\mathrm{\sigma}_{\footnotesize\pp} / \mathrm{d}\eta \mathrm{d}\pT$ is differential charged-particle production cross section in \pp collisions.

This analysis~\cite{atlas_raa} uses 0.15\,nb${}^{-1}$ of \PbPb data collected with the ATLAS detector~\cite{Aad:2008zzm} in 2010 and 2011 at $\sqn=2.76$\,\TeV, and 4.2\,pb${}^{-1}$ of \pp data at the same value of $\sqs$ recorded in 2011 and 2013. The events were taken either with minimum bias triggers or with jet triggers. The jet reconstruction used energy deposits in the calorimeter. The jet triggers used \akt algorithm~\cite{anti-kt} with the radius parameter of $R=0.2$ and 0.4 for \PbPb and \pp collisions, respectively.In each data-taking period, the triggers were chosen to take as much advantage of the instantaneous luminosity as possible.

Monte Carlo (MC) simulations were used to correct the detector-level charged-track spectra to the particle-level. Hard-scattering processes for \PbPb collisions were produced using PYTHIA event generator~\cite{Pythia}. To correct \PbPb data sample from 2010, hard-scattering events were embedded into events produced by HIJING event generator~\cite{Wang:1991hta}. To correct \PbPb data sample from 2011, real \PbPb events recorded for this purpose were used as underlying events. The \pp simulated events were generated by PYTHIA for both 2011 and 2013. The MC samples for all data-takings were produced in different exclusive kinematic intervals of outgoing partons in the \mbox{$2\rightarrow2$} hard scattering process, allowing sufficient statistics over wide range of track $\pT$. For the 2013 MC samples, the kinematic intervals were defined in terms of leading jet $\pT$, otherwise the approach was analogous.

The event centrality is estimated using total transverse energy measured by the forward calorimeter in the range $3.1<|\eta|<4.9$. The $\avgTaa$ values are estimated from the Glauber model~\cite{Glauber} for each of the studied centrality intervals.

Charged-particle tracks are reconstructed in the ATLAS inner detector over pseudorapidity region $|\eta|<2.5$ and over full azimuth angle, with the minimum $\pT$ of 0.5\,\GeV. Tracks are measured using combination of silicon pixel detector (Pixel), silicon microstrip detector (SCT), and a straw tube transition ration tracker (TRT). Whole system rests in a 2\,T magnetic field.

The tracks reconstructed in \PbPb collisions are required to have at least two hits in the Pixel, one being in the innermost layer, and at least seven hits in the SCT. Tracks cannot have any Pixel or SCT holes (a hole is an absence of a hit predicted by the track trajectory). To ensure that tracks originate from the event vertex of the collision and that they are not fake tracks (spurious combination of hits) or secondary tracks (decay products of long-lived particles such as K${}_\mathrm{S}$ or $\Lambda$), tracks are required to have ratios of $\big| \dz/\edz \big|$ and $\big|\zz/\ezz \big|$ less than three (track parameters $\dz$ and $z_0$ are distances of the closest approach of the track to the event vertex in the transverse and longitudinal directions, respectively; $\theta$ is an angle between track and z-axis; $\sigma$ stands for an uncertainty of a variable in its subscript). 

Generally, requirements imposed on the tracks in \pp collisions are looser. The tracks are required to have at least one Pixel hit, at least six SCT hits and at least eight TRT hits (only for $\pT>6$\,\GeV). A hit in the Pixel innermost layer is required only if such hit is expected from the track trajectory. Tracks are allowed to have one SCT hole; no Pixel holes are allowed. The TRT hit requirement limits the analysis to $|\eta|<2$, which is the coverage of the TRT detector. Due to lower multiplicity in \pp collisions, track reconstruction is not as challenging as in case of \PbPb collisions. Also, the vertex is not defined so well as in \PbPb collisions. Thus, only a loose requirement of $\left|\dz\right|<1.5$\,mm is applied. 

Tracks with high \pT are further required to be matched to \akt jets with distance parameter $R=0.2$ and $R=0.4$ in \PbPb and \pp collisions, respectively.

Contribution of leptons, usually coming from electroweak decays, is excluded from the results, as leptons follow different scaling than hadrons.

The data are corrected in order to remove detector effects and reconstruction biases. The corrections are derived from data or MC simulations. The track spectra are corrected for jet trigger efficiency and for vertex reconstruction efficiency. These corrections are relatively small and do not exceed a few per cent. Further, the track spectra are corrected for fake and secondary tracks, for the momentum resolution and for the track reconstruction efficiency. These corrections have more serious impact on the results. They are functions of \pT, $|\eta|$ and centrality. Correction for fake and secondary tracks is up to 10\% at large $|\eta|$ and low $\pT$. It rapidly decreases with increasing $\pT$ and at $\pT \gtrsim 4$\,\GeV, the correction is below 1\% for all samples. At $\pT \gtrsim 70$\,\GeV, the correction becomes more important again and it reaches up to 6\% at the highest $\pT$. Track momentum resolution is corrected with an iterative Bayesian unfolding procedure~\cite{BayesUnf}. Results are obtained with two iterations. The track reconstruction efficiency is estimated from the MC simulation after excluding fake and secondary tracks from consideration. The efficiencies are smoothed by fitting to minimize the effect of statistical fluctuations at high $\pT$. An example of efficiencies is shown in figure~\ref{fig1}.

\begin{Figure}
	\centering
	\mbox{ \hspace*{-0.4cm} \includegraphics[width=\linewidth]{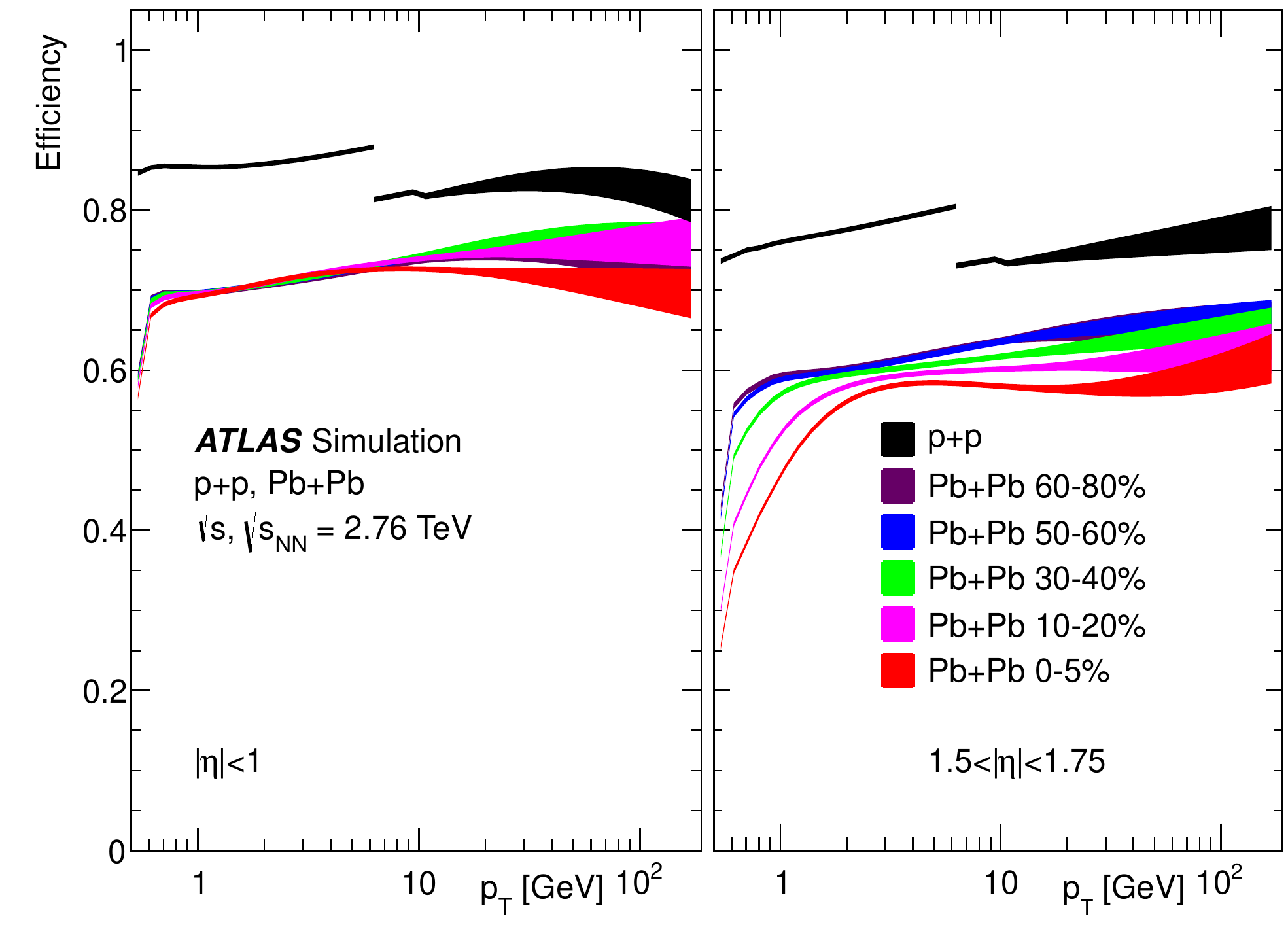}}
	\captionof{figure}{Track reconstruction efficiency for five centrality classes of \PbPb collisions as well as for \pp collisions~\protect\cite{atlas_raa}. The discontinuity of the \pp efficiency is caused by the change of the TRT hit requirement. The width of the bands represent the systematic uncertainties.}
	\label{fig1}
\end{Figure}

Systematic uncertainties are evaluated by varying parameters in the analysis for each source and observing the effects on the yields. For uncertainties of $\Raa$, variations are applied simultaneously in \PbPb and \pp samples, so the final uncertainties reflect the correlations between numerator and denominator. Main contributions come from the uncertainty of fake tracks at high \pT, reaching up to 20\%, and from possible differences in the distortion of the track $\pT$ in MC simulations and data, also reaching up to 20\%. The uncertainty from the calculation of $\avgTaa$ is at most 13\%. Variation of the track selection requirements introduces an uncertainty no more than 10\%. Uncertainty associated with the unfolding procedure is at most 8\%. Other sources of the uncertainties, such as detector material or track reconstruction efficiency, do not exceed 6\%.

\section{Results}

The corrected charged-hadron spectra measured in \PbPb collisions at $\sqn=2.76$\,\TeV are shown in figure~\protect\ref{fig2}. Going from peripheral to central events, the \mbox {$\Taa$-scaled} \PbPb yields increasingly deviate from the \pp spectra.

\begin{Figure}
	\centering
	\includegraphics[width=0.98\linewidth]{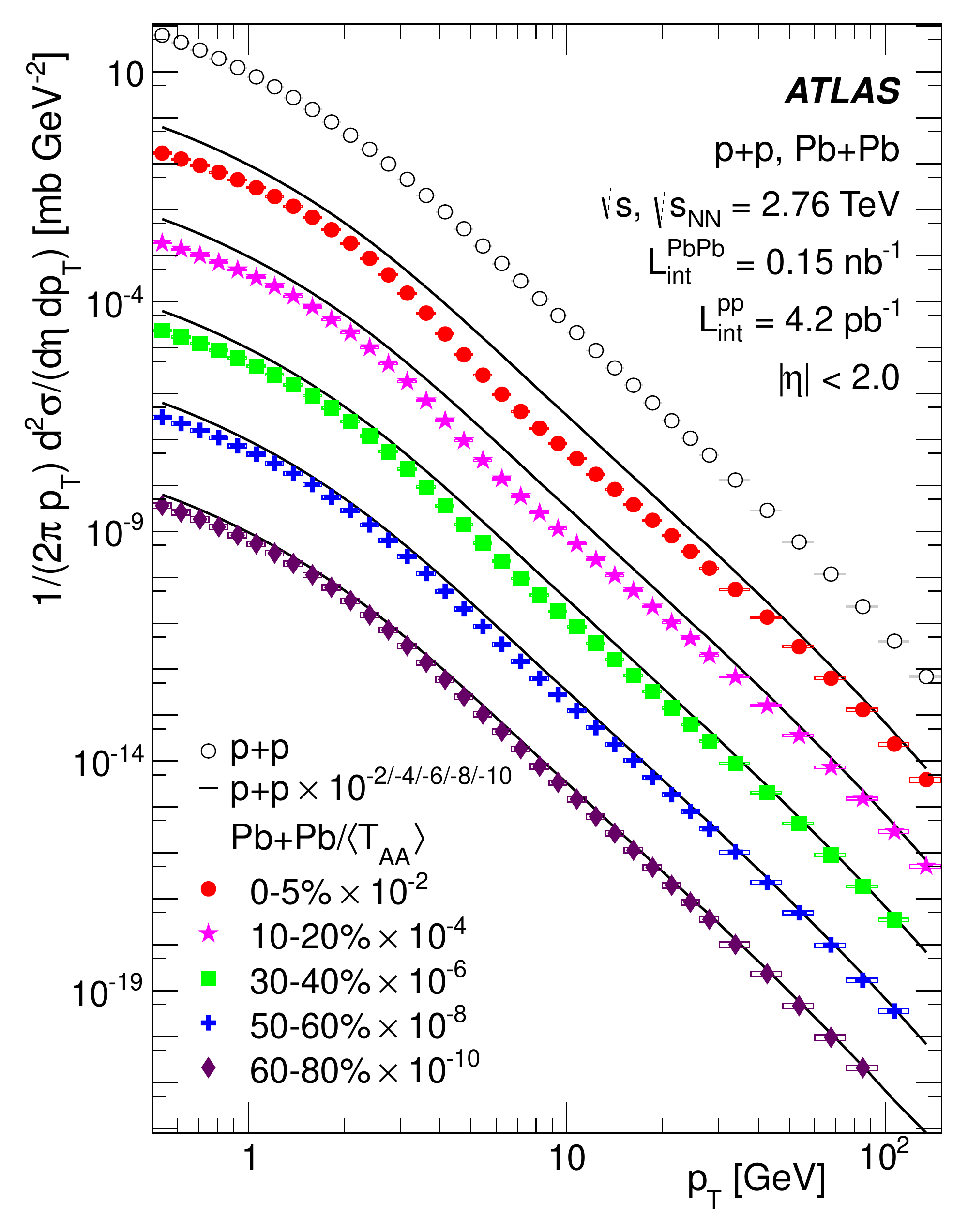}
	\captionof{figure}{Fully corrected charged-hadron spectra~\protect\cite{atlas_raa}. Spectra from \PbPb collisions are scaled down by powers of ten and plotted together with the \pp cross section scaled by the same factor with solid line. Statistical uncertainties are smaller than the marker size. Systematic uncertainties are shown by open boxes. }
	\label{fig2}
\end{Figure}

Figure~\ref{fig3} shows the nuclear modification factor $\Raa$ as a function of $\pT$. It shows a characteristic shape which becomes more pronounced for more central collisions. It first increases and reaches a maximum at $\pT \approx 2$\,\GeV. Then it decreases to a minimum at $\pT \approx 7$\,\GeV, where the suppression is the strongest. Above this $\pT$, $\Raa$ increases up to $\pT \approx 60$\,\GeV and then reaches a plateau. The plateau is consistent with zero slope. In the most central collisions, it has a value of \mbox{$0.55\pm0.01(stat.)\pm0.04(syst.)$}.

Figure~\ref{fig4} shows the nuclear modification factor $\Raa$ as a function of $\eta$ in four momentum intervals corresponding to the local maximum ($1.7<\pT<2.0$\,\GeV), the local minimum (\mbox{$6.7<\pT<7.7$\,\GeV}), the plateau region (\mbox{$59.8<\pT<94.8$\,\GeV}) and to the $\pT$ interval where $\Raa$ has an intermediate value (\mbox{$19.9<\pT<22.8$\,\GeV}). For all intervals $\Raa$ shows a weak dependence on $|\eta|$ and it is generally consistent with a flat behaviour within the statistical and systematic uncertainties.

\begin{Figure}
	\centering
	\mbox{ \hspace*{-0.69cm} \includegraphics[width=1.04\linewidth]{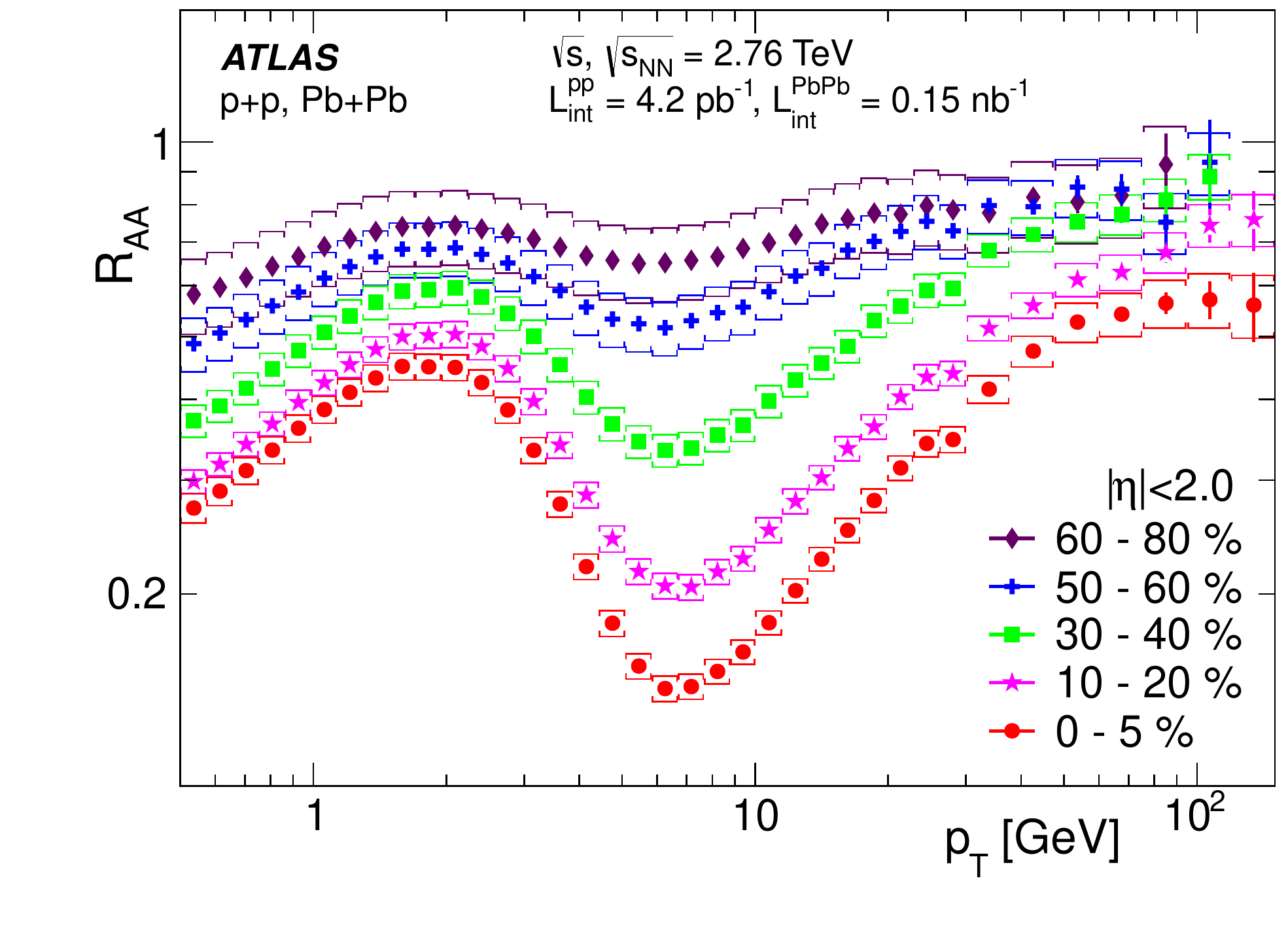}}
	\captionof{figure}{The nuclear modification factor $\Raa$ as a function of $\pT$~\protect\cite{atlas_raa}. Statistical uncertainties are shown with vertical bars and systematic uncertainties with brackets.}
	\label{fig3}
\end{Figure}

\begin{Figure}
	\centering
	\includegraphics[width=0.98\linewidth]{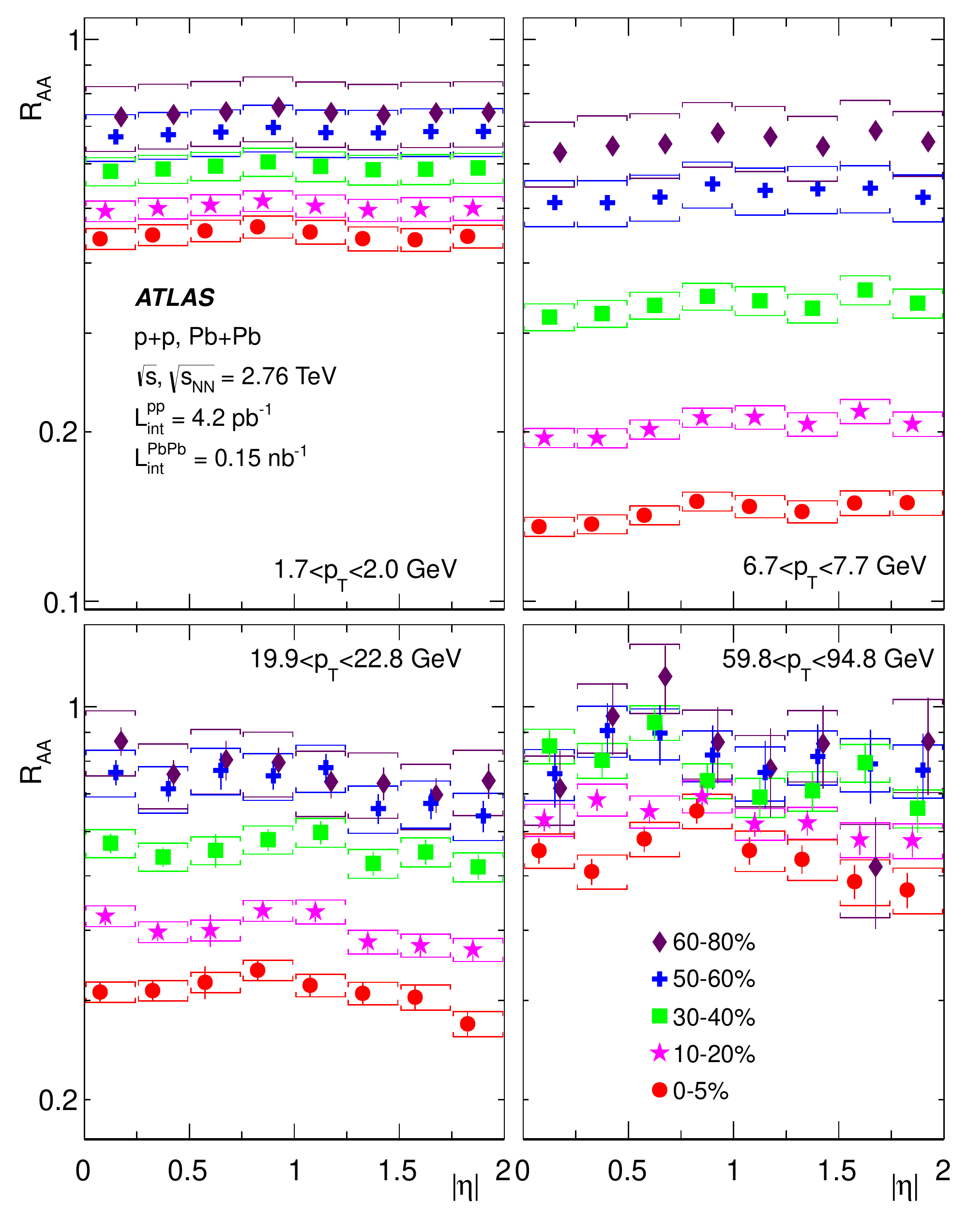}
	\captionof{figure}{The nuclear modification factor $\Raa$ as a function of $\eta$ in four momentum intervals discussed in the text~\protect\cite{atlas_raa}. Statistical uncertainties are shown with vertical bars and systematic uncertainties with brackets.}
	\label{fig4}	
\end{Figure}

\vspace{0.1cm}

Figure~\ref{fig5} shows a comparison of $\Raa$ measured for charged hadrons and that for jets. The charged-hadron $\Raa$ is shown for the plateau region. Jets are shown for three different $\pT$ regions. The lowest one agrees to the $\pT$ region of charged hadrons. Jets from the intermediate intervals produce main portion of shown charged hadrons. Having fixed $\pT$ of composing hadrons, two contradictory trends meet here -- the decreasing $\pT$ spectrum of the jets, not unlike the charged-hadron spectrum, and the increasing fragmentation function when going toward lower $z$. Jets from the highest $\pT$ region are composed mostly from shown hadrons, due to the fragmentation functions. It is the $\Raa$ of jets from the highest $\pT$ interval that overlap with hadron $\Raa$ the best, at least for more central collisions. 

\begin{Figure}
	\centering
	\mbox{ \hspace*{-0.69cm} \includegraphics[width=1.04\linewidth]{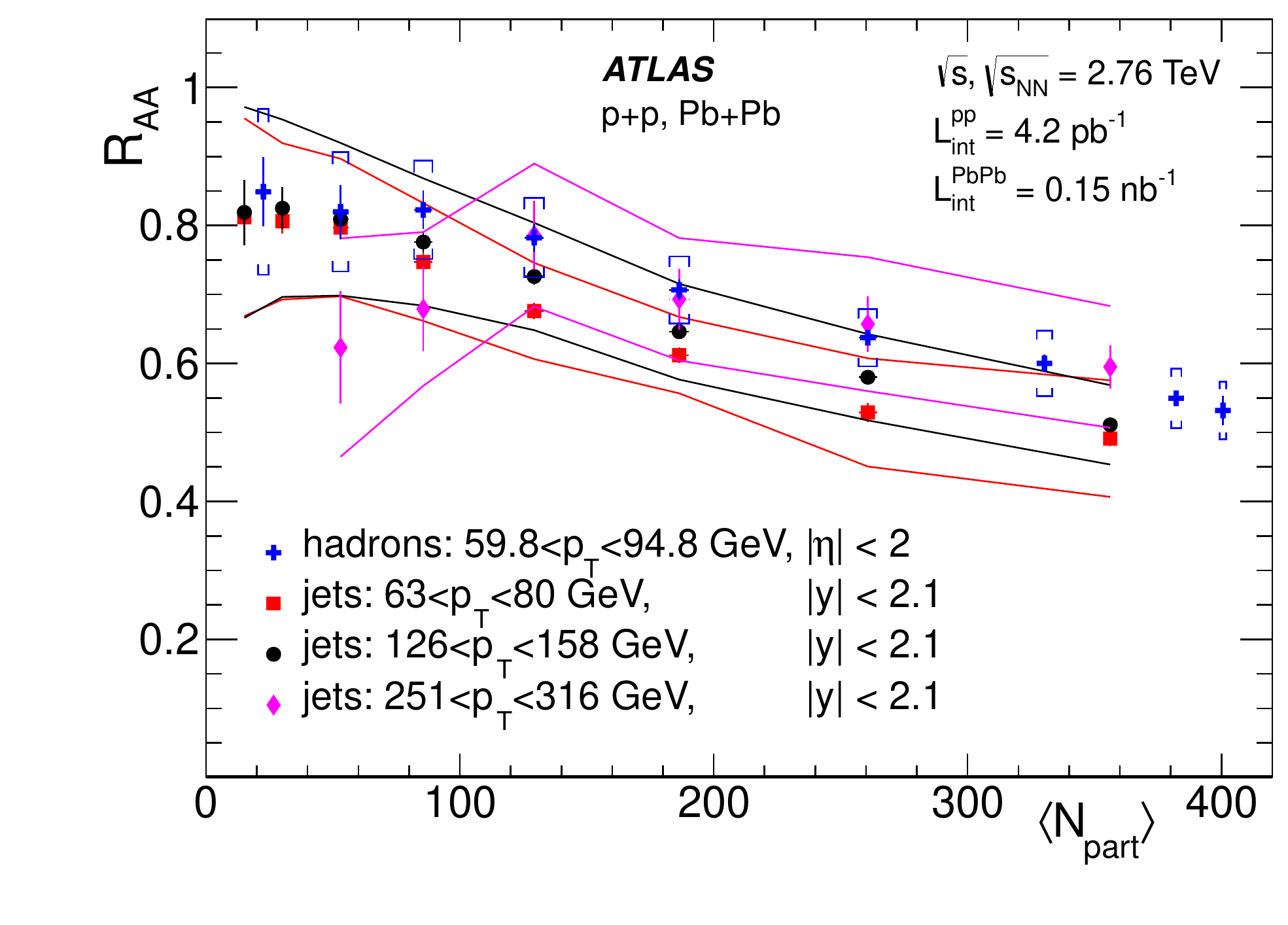}}
	\captionof{figure}{The nuclear modification factor $\Raa$ as a function of $\avgNpart$ measured for jets~\protect\cite{atlas_jet_raa} and charged hadrons~\protect\cite{atlas_raa,atlas_raa_aux}. Statistical uncertainties are shown with vertical bars and systematic uncertainties with brackets (charged hadrons) or bands (jets).}
	\label{fig5}		
\end{Figure}

\section*{Acknowledgements}
The speaker is supported by Charles University in Prague (projects PRVOUK P45 and UNCE 204020/2012) and by M\v{S}MT \v{C}R (project INGO LG13009).

\bibliographystyle{elsarticle-num}
\bibliography{Balek_P}

\begin{thebibliography}{10}
\expandafter\ifx\csname url\endcsname\relax
  \def\url#1{\texttt{#1}}\fi
\expandafter\ifx\csname urlprefix\endcsname\relax\def\urlprefix{URL }\fi
\expandafter\ifx\csname href\endcsname\relax
  \def\href#1#2{#2} \def\path#1{#1}\fi

\bibitem{Aj_atlas}
{ATLAS Collaboration}, Phys. Rev. Lett. 105 (2010) 252303.
\newblock \href {http://arxiv.org/abs/1011.6182} {\path{arXiv:1011.6182}}.

\bibitem{Aj_cms}
{CMS Collaboration}, Phys. Rev. C 84 (2011) 024906.
\newblock \href {http://arxiv.org/abs/1102.1957} {\path{arXiv:1102.1957}}.

\bibitem{atlas_jet_raa}
{ATLAS Collaboration}, Phys. Rev. Lett. 114~(7) (2015) 072302.
\newblock \href {http://arxiv.org/abs/1411.2357} {\path{arXiv:1411.2357}}.

\bibitem{alice_jet_raa}
{ALICE Collaboration, J. Adam et al.}, Phys. Lett. B 746 (2015) 1--14.
\newblock \href {http://arxiv.org/abs/1502.01689} {\path{arXiv:1502.01689}}.

\bibitem{frag_atlas}
{ATLAS Collaboration}, Phys. Lett. B 739 (2014) 320--342.
\newblock \href {http://arxiv.org/abs/1406.2979} {\path{arXiv:1406.2979}}.

\bibitem{frag_cms}
{CMS Collaboration}, Phys. Rev. C 90 (2014) 024908.
\newblock \href {http://arxiv.org/abs/1406.0932} {\path{arXiv:1406.0932}}.

\bibitem{atlas_raa}
{ATLAS Collaboration}, JHEP 09 (2015) 050.
\newblock \href {http://arxiv.org/abs/1504.04337} {\path{arXiv:1504.04337}}.

\bibitem{cms_raa}
{CMS Collaboration}, Eur. Phys. J. C 72 (2012) 1945.
\newblock \href {http://arxiv.org/abs/1202.2554} {\path{arXiv:1202.2554}}.

\bibitem{alice_raa}
{ALICE Collaboration, B. Abelev et al.}, Phys. Lett. B 720 (2013) 52--62.
\newblock \href {http://arxiv.org/abs/1208.2711} {\path{arXiv:1208.2711}}.

\bibitem{brahms}
{BRAHMS Collaboration, I. Arsene et al.}, Nucl. Phys. A 757 (2005) 1--27.
\newblock \href {http://arxiv.org/abs/nucl-ex/0410020}
  {\path{arXiv:nucl-ex/0410020}}.

\bibitem{phobos}
{PHOBOS Collaboration, B.B. Back et al.}, Nucl. Phys. A 757 (2005) 28--101.
\newblock \href {http://arxiv.org/abs/nucl-ex/0410022}
  {\path{arXiv:nucl-ex/0410022}}.

\bibitem{star}
{STAR Collaboration, J. Adams et al.}, Nucl. Phys. A 757 (2005) 102--183.
\newblock \href {http://arxiv.org/abs/nucl-ex/0501009}
  {\path{arXiv:nucl-ex/0501009}}.

\bibitem{phenix}
{PHENIX Collaboration, K. Adcox et al.}, Nucl. Phys. A 757 (2005) 184--283.
\newblock \href {http://arxiv.org/abs/nucl-ex/0410003}
  {\path{arXiv:nucl-ex/0410003}}.

\bibitem{Aad:2008zzm}
{ATLAS Collaboration}, JINST 3 (2008) S08003.

\bibitem{anti-kt}
M.~Cacciari, G.~P. Salam, G.~Soyez, JHEP 0804 (2008) 063.
\newblock \href {http://arxiv.org/abs/0802.1189} {\path{arXiv:0802.1189}}.

\bibitem{Pythia}
T.~Sjostrand, S.~Mrenna, P.~Skands, JHEP 0605 (2006) 026.
\newblock \href {http://arxiv.org/abs/hep-ph/0603175}
  {\path{arXiv:hep-ph/0603175}}.

\bibitem{Wang:1991hta}
X.-N. Wang, M.~Gyulassy, Phys. Rev. D 44 (1991) 3501--3516.

\bibitem{Glauber}
M.~L. Miller, K.~Reygers, S.~J. Sanders, P.~Steinberg, Ann. Rev. Nucl. Part.
  Sci. 57 (2007) 205.
\newblock \href {http://arxiv.org/abs/nucl-ex/0701025}
  {\path{arXiv:nucl-ex/0701025}}.

\bibitem{BayesUnf}
G.~D'Agostini, Nucl. Instrum. Meth. A 362 (1995) 487--498.

\bibitem{atlas_raa_aux}
{ATLAS Collaboration}, \url{http://atlas.web.cern.ch/
  Atlas/GROUPS/PHYSICS/PAPERS/HION-2011-03}.

\end{thebibliography}

\end{multicols}

\end{document}